\documentclass[]{JHEP3}
\usepackage{graphicx}

\newcommand{\be}{\begin{equation}}
\newcommand{\ee}{\end{equation}}
\newcommand{\bea}{\begin{eqnarray}}
\newcommand{\eea}{\end{eqnarray}}

\def\jb{\bar{j}}
\def\Wb{{\bar{W}}}
\def\zb{{\bar{z}}}
\def\phib{{\bar{\phi}}}

\title{A cigar-like Universe}
\author{B.~de Carlos \\ Department of Physics and Astronomy, University of Sussex, \\ Falmer, Brighton BN1 
9QJ, UK  \\
CERN, TH Division, CH-1211 Geneva 23, Switzerland\\ E-mail: \email{B.de-Carlos@sussex.ac.uk}} 
\author{J.M.~Moreno \\ Instituto de Estructura de la Materia, CSIC, Serrano 123, 28006 Madrid, Spain \\
IFT C-XVI, UAM, Cantoblanco 28049 Madrid, Spain \\ 
E-mail: \email{jesus@makoki.iem.csic.es}} 
\abstract{We study the localization of gravity on a string-like topological defect within a 6-dimensional
space-time. Assuming zero cosmological constant we find complete numerical solutions 
to a set of first-order, Bogomol'nyi-Prasad-Sommeferld (BPS)-like, equations for the metric and the scalar field,
where the dynamics of the latter are dictated by a supergravity-type potential.
Our axially symmetric solutions have no deficit angle and factorize as $AdS_5 \times S_1 $ 
far from the core. They are regular everywhere, providing  complete smooth cigar-like geometries. 
The total energy of these configurations depends only on the boundary conditions for the warp factor
and it is shown to vanish.}
\preprint{SUSX-TH-03-011 \\
CERN--TH/2003--218\\
IEM-FT/232-03\\
IFT-UAM-CSIC-03-36}

\begin{document}

\section{Introduction}
Already since the early eighties \cite{Akama:jy,Rubakov:1983bz,Visser:qm} physicists have toyed with the idea 
of considering our 4D Universe as the internal space of a topological defect living in higher dimensional space-time. 
Such defect could either be a domain wall (if the total of space-time dimensions is equal to five), a global string 
(in six dimensions), a monopole (in seven dimensions) or an instanton (in eight dimensions). This
idea received further boosts, first with the realization that solitons in string theory (i.e. D-branes) were
ideal candidates to localize gauge and matter fields \cite{Polchinski:1995mt}, followed by the work of Randall and 
Sundrum \cite{Randall:1999ee} on the localization of gravity in a three-brane domain wall in five space-time dimensions.

Since then exhaustive work has been devoted to the study of five-dimensional models and slightly less to higher
dimensional setups. In particular, for D=6 or the global string case, a number of issues have already been 
discussed. Whereas in D=4 the global string in the presence of a static metric is singular \cite{Cohen:1988sg}, and
finite when we include time-dependence \cite{Gregory:1996dd}, we see that the corresponding D=6 model continues to be 
singular when the metric is time-independent \cite{Cohen:1999ia} and in the presence of a zero cosmological constant, 
while it becomes regular again if a negative cosmological constant is introduced \cite{Gregory:1999gv}. So far we are 
always assuming a system with gravity and a scalar field with a U(1) invariant interaction. Let us mention that, 
already, a fair amount of work has been devoted to the study of the local string in D=6 \cite{Gherghetta:2000qi} 
and, in particular, to finding numerical solutions to the Abelian Higgs model \cite{Giovannini:2001hh}.

Also, a very important aspect of these D=6 models is their potential phenomenological applicability. In that respect
the localization of matter fields has been studied in 
refs.~\cite{Oda:2000zj}-\cite{Midodashvili:2003ib}.
Other issues, such as their cosmology~\cite{Cline:2003ak} or the status of the cosmological constant 
problem~\cite{Wetterich:1984rv}-\cite{Nilles:2003km}
have also been considered elsewhere.

In this letter we are going to consider a class of models where gravity interacts with a complex scalar field whose
dynamics are dictated by a supergravity-like structure. Moreover, we shall restrict 
ourselves to solving a set of first-order differential equations, the BPS equations, which can be deduced from the 
full, second-order, set of Einstein equations plus the equation of  motion for the scalar field. Our goal is,
given a K\"ahler potential and a superpotential for our scalar field, to find full numerical solutions for the field 
configuration and warp factors of the metric which localize gravity and give rise to an acceptable hierarchy
between the Planck and electroweak scales in D=4. In addition, and based on our numerical results, we aim to establish 
general conditions for the existence of such trapping solutions.

In the next section we present the models we want to study and the set of differential equations we want to solve.
We also draw general conclusions about the structure of the solutions we are after. In section~3 we present 
models where the scalar field has canonical kinetic terms, solve the equations of motion and explain why the 
solutions obtained are not satisfactory. In section~4 we turn to models with a more complicated K\"ahler potential
for which there are acceptable analytic solutions, and we explain their general characteristics. Finally, 
in section~5, we draw general results on the structure of these solutions and present the main conclusions.


\section{The model: gravity and a scalar field}

From now on we shall essentially follow the notation of ref.~\cite{Carroll:1999mu}, from where 
we also borrow their derivation of first-order BPS equations out of the second order Einstein
plus field equations. We start off with an action in D-dimensions given by
\be
S = \int d^Dx\, \sqrt{| g|}  \left({1\over 2} R -  K_{i\jb} g^{ab}
(\partial_a \phi^i)  (\partial_b \bar\phi^j) - V(\phi,\bar\phi)\right) \ ,
\label{lagrangian}
\ee
where we work in units where $8 \pi G_D=1$, $\phi_i$ are complex scalar fields, 
$K_{i\jb}=\partial^2 K/\partial \phi_i \partial \bar{\phi}_j$ is a K\"ahler metric, 
with $K$ the K\"ahler potential. 
We are going to consider from now on one scalar field $\phi$ whose potential is given by the following 
expression
\be
V = \left(\frac{D-2}{4}\right) {\rm e}^K \left[ (D-2)
K^{\phi\bar{\phi}} W_{;\phi} \Wb_{;\phib} - 2(D-1) W \Wb \right],
\label{potential}
\ee
where $W(\phi)$ is a holomorphic function and plays the role of a superpotential. Moreover
\be
W_{;\phi} \equiv {{\partial W}\over{\partial\phi}} +  {{\partial K}\over{\partial\phi}}W\ .
\ee
This form of the potential given by eq.~(\ref{potential}) 
guarantees, at least for canonical K\"ahler metrics,
vacuum stability with respect to small fluctuations of the $AdS_D$ geometry, see ref.~\cite{Townsend:iu}.

Once we have defined the matter content of our model, we can introduce the gravity sector. 
Since we are looking for string-like topological defects that could localize gravity, leading to
our known four-dimensional world near the core of the string, we will fix from now on D=6.
A general metric consistent with four-dimensional Poincar\'e invariance is given by
\be
ds^2 = {\rm e}^{2A(z,\zb)}\eta_{\mu\nu}dx^\mu dx^\nu
  + {1\over 2} {\rm e}^{2B(z,\zb)}\left(dzd\zb + d\zb dz\right)\ ,  
\label{metric}
\ee
where $\mu,\nu = 1,...4$, $z  \equiv  x^5 + ix^6$, the signature of
the D=4 metric is mostly plus and the two warp factors, $A$ and $B$, 
are functions of the extra coordinates $z$, $\bar{z}$. As we mentioned in the introduction, 
it is our intention to solve a reduced first-order version of the Einstein and scalar field equations, 
as derived in ref.~\cite{Carroll:1999mu}, which is given by 
\bea
\partial_z \phi & = &  
{\rm e}^{N^*} K^{\phi\bar{\phi}} \Wb_{;\bar{\phi}} \ ,  \label{eqphi} \\
\partial_z  A & = & - {1\over 2}
{\rm e}^{N^*} \Wb \ , \label{eqA} \\
\partial_z N & =  & K_{\phi} \partial_z \phi +     \partial_z  A
 = {1\over 2} {\rm e}^{N^*} (2 K^{\phi \bar{\phi}} K_{\phi} \Wb_{;\bar{\phi}} - \Wb) \ , 
\label{eqN}
\eea
where the function $N$ is defined as
\be 
N\equiv B+\frac{K}{2}+iJ \;, 
\label{N}
\ee
with $J$ a real function. Note that equation (\ref{eqA}) can be integrated once 
the system (\ref{eqphi}, \ref{eqN}) has been solved.

We are interested in axially symmetric solutions, therefore we will assume that, 
under a rotation 
$z \rightarrow {\rm e}^{i \theta } z$, the scalar field transforms as $\phi  \rightarrow  {\rm e}^{i q \theta } \phi$,
the K\"ahler potential is invariant and $J$ gets shifted $ J \rightarrow  J +  n \theta$  for some integers $q, n$. 
Notice that this implies that $W$ has to be a monomial,  $W = \phi^p$, and that
\be
n + p q = 1 \;.
\label{enteros}
\ee
In order to solve the system of equations give above we are going to parametrize the relevant functions as follows
\bea    
\phi(z, \zb)      & = &   f(r)  {\rm e}^{iq\theta + i\theta_\phi } \; , \label{deff} \\
{\rm e}^{2 A(z, \zb)}   & = &   g(r)  \; ,  \label{defg} \\
{\rm e}^{N(z, \zb)}     & = &   h(r)  {\rm e}^{ij(r^2) + in\theta + i\theta_N} \; , \label{defh} 
\eea
where $r^2 = z \zb$, and we have implicitly assumed that $J(z,\zb)=j(z\zb)+n\theta$. For the time being we are
going to set $j(z\zb)=0$, although we shall return to this point later on. 
We want the functions $f$, $g$ and $h$ to be real and, for phenomenological purposes, we
would like $g$ to be positive-definite. Therefore, from eq.~(\ref{eqA}) we obtain a further constraint, namely
\be
\theta_N + p \theta_\phi = 0 \;,
\ee
modulo $\pi $. We can always rotate $\theta_{\phi}$ away and, in such basis, $\theta_N=0$. This is the choice
we shall make from now on.

\section{Canonical kinetic terms}

We are going to classify the models we want to study according to their K\"ahler potential. The simplest case is,
then, that of considering canonical kinetic terms, i.e.  $K(\phi, \phib) = \phi \phib$. The scalar potential,
eq.~(\ref{potential}), is given by
\be
V ( \phi, \phib) = {\rm e}^{ \phi \phib } 2 (\phi \phib)^{p-1} 
[ 2 (p + (\phi \phib))^2 - 5 \phi \phib] \;.
\label{potcanp}
\ee        
The system of BPS-equations becomes
\bea
\label{bpsf}
 \frac{d f}{d r}  + \frac{q}{r}f & = & 2 h  f^{p-1} (p+f^2) \;, \\
\label{bpsh}
 \frac{d h}{d r} + \frac{n}{r}h & = & 2 h^2  f^p ( 2  (p+f^2) - 1) \;,  \\
\label{bpsg}
  \frac{d g }{d r} & = & - 2 g h f^p \;.
\eea
As we have already mentioned, the equation for $g$ can be immediately integrated once the two first ones 
have been solved. Note that these equations are invariant under the transformations 
$f \rightarrow - f$,  $h \rightarrow  (-1)^{|p|} \, h$. Since the transverse metric is proportional $h^2(r)$, 
this function could be either positive or negative but, in any case, has to be different from zero.

The condition for trapping gravity is given by
\be
\int dx^4 dx^5 g^{00} \sqrt{|g|}
\sim
\int dr \; {\rm e}^{2 A + 2 B } r
=
\int dr \; g(r) h^2(r)  {\rm e}^{-f^2(r)} r  < \infty \;\;.
\label{criterio}
\ee
If we assume that the solution is regular everywhere (including the origin) 
the condition for a finite Planck mass depends on the behaviour 
of the configuration for large $r$ values.

It is easy to check that a possible consistent ansatz in the large $r$ region is
\bea
f(r)|_{r \rightarrow \infty}  & \sim & f_0 \;\;, \\
h(r)|_{r \rightarrow \infty}  & \sim & \frac{h_0}{r} \;\;, 
\label{infty}
\eea 
with $f_0, h_0$ constants. From  eq.~(\ref{bpsf}) we can calculate
\be
h_0  =   \frac{q}{2} \frac{f_0^{2-p}}{f_0^2 + p} \;\;.
\label{h0}
\ee
We can also calculate $g$ using this ansatz. The result is
\be
g(r)|_{r \rightarrow \infty}  \sim g_0 \, r^{- 2 h_0  f_0^p} \;\;.
\label{g0}
\ee
Then, the integral given by eq.~(\ref{criterio}) is convergent when 
\be
h_0  \, f_0^p > 0 \;\;, 
\label{trapping}
\ee
i.e. when
\be
\frac{q}{2} \frac{f_0^2}{f_0^2 + p} > 0
\label{trapalt}
\ee
is fulfilled. 

As mentioned above, the validity of this ansatz relies on the existence of a real solution to the 
algebraic equation for $f_0$. Combining eqs.~(\ref{bpsf}, \ref{bpsh}) we get
\be
f_0^2  =   \frac{1}{4} \left( 1 - 3 p \pm \sqrt{1 - 6 p + p^2 }\right) \;\;.
\label{f0}
\ee
In the preceding discussion, the value of $p$, the degree of the superpotential term, was an arbitrary 
integer. From eq.~(\ref{f0}) we deduce that a real value for $f_0$ implies $p \leq 0$. However, a non singular 
scalar potential implies $p \geq 1$ or $ p = 0$; therefore we will fix $ p = 0$. 
Note that the trapping condition, eq.~(\ref{trapalt}), for that case translates into $q > 0$. 

After studying the necessary conditions to have gravity trapping in the large $r$ region, we 
should check the hypothesis of regularity at $r=0$. Once we fix $W = 1$ it turns out that there 
are no regular, consistent, boundary  conditions at the origin which are compatible with this choice. 
This can be also inferred from the shape of the potential 
\be
V ( \phi, \phib)|_{p = 0}  = {\rm e}^{ \phi \phib } 2  ( 2 \phi \phib - 5 ) \;\;,
\label{potcan}
\ee        
shown in fig.~\ref{fig:Vp}. There we can see that, at least up to the minimum, the value of the potential
energy is negative. This is a crucial fact that prevents the model from working and which we shall
explain in detail in section~5.
\FIGURE{\centerline{\includegraphics*[width=10cm, draft=false]{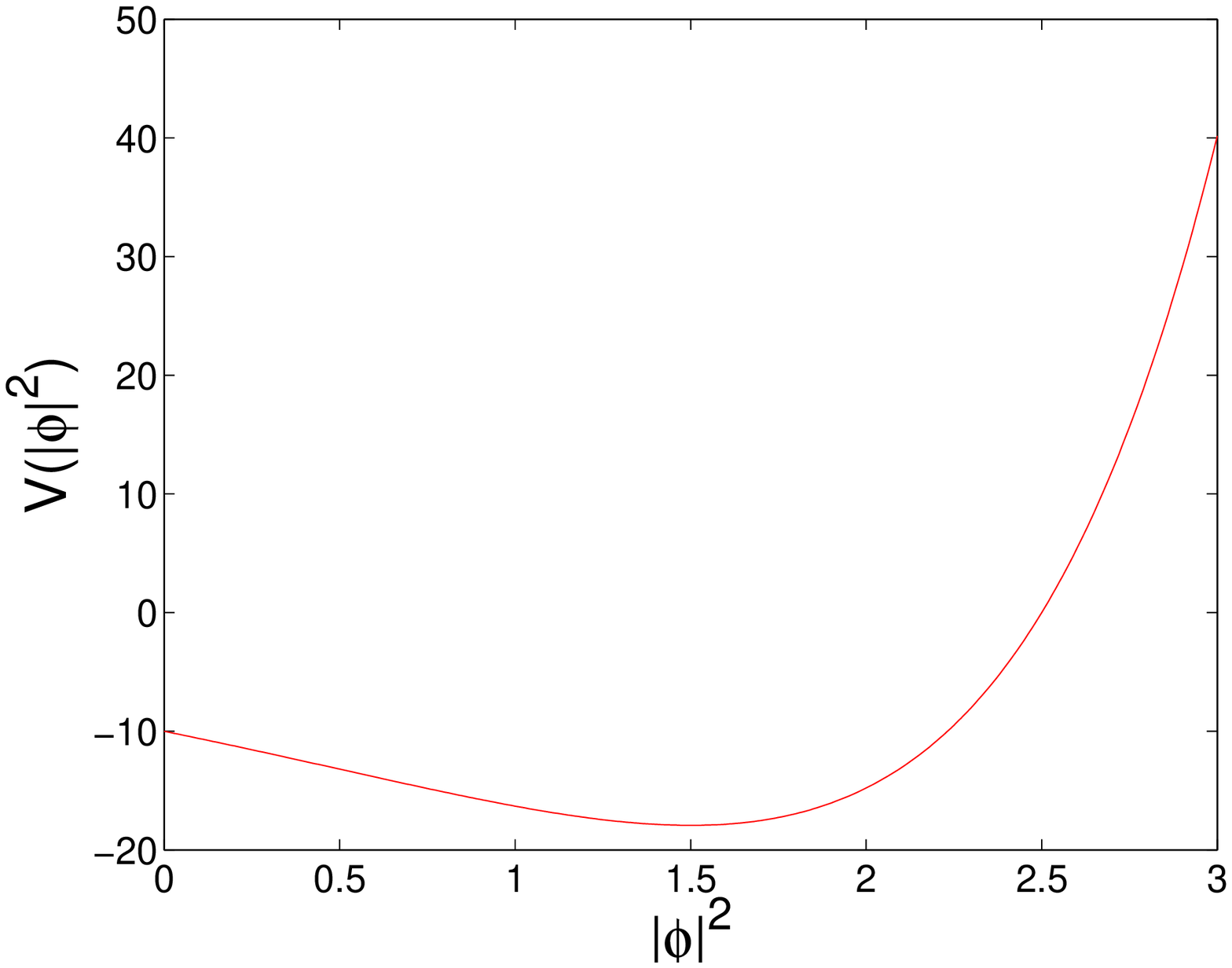}}
\caption{Scalar potential, as given in eq.~(\ref{potcan}), as a function of 
$|\phi|^2$.}
\label{fig:Vp}
}
%

\section{Non-canonical kinetic terms}

Motivated by the previous discussion, we have generalized the analysis to the case where the K\"ahler 
potential is not canonical. The BPS equations for this case are
\begin{eqnarray}    
 \frac{d f}{d r} + \frac{q}{r}f & = & 2 h  f^{p-1} 
\frac{p+f^2 k'}{k'+ f^2 k''} \;, \\
  \frac{d h}{d r} + \frac{n}{r}h & = & 
   h^2  f^p ( 2 k' \frac{p+f^2 k'}{k'+ f^2 k''}  - 1 )  \;, \\
  \frac{d g}{d r} & = & - 2 g h f^p \;, 
\end{eqnarray}
where $ k(\phi \phib) = K(\phi, \phib)$ and $'$ denotes a derivative with respect to $\phi \phib$. The 
corresponding equations~(\ref{bpsf}-\ref{bpsg}) for the canonical case are recovered by taking $k' = 1, k'' = 0$.

The equivalent to equations~(\ref{h0}) and~(\ref{f0}) for the asymptotic ansatz are given by
\begin{eqnarray}    
   h_0 & = & \frac{q}{2}f_0^{2-p} 
\frac {k'_0+ f_0^2 k''_0} {p+f_0^2 k'_0} \;, \\
 \frac{1}{n-1} ( 2 k'_0 \frac {k'_0+ f^2 k''_0} {p+f_0^2 k'_0} -1) 
 & = & 
  2 f_0^{-2} 
\frac {k'_0+ f_0^2 k''_0} {p+f_0^2 k'_0} \;, 
\end{eqnarray}
where $k'_0=  k'( f_0^2)$, $k''_0 =  k''( f_0^2)$.

For a regular K\"ahler potential, the trapping condition is the same as the one for the canonical case, 
eq.~(\ref{trapping}), namely $h_0 f_0^p > 0 $ which, in terms of $f_0$, reads
\be
q \frac{f_0^2}{2} 
\frac {k'_0+ f_0^2 k''_0} {p+f_0^2 k'_0} > 0 \;\;.
\ee
Once we have determined all possible phenomenological constraints on the model, we should try to look for solutions
of the BPS equations that fulfill them. Although we have not performed a systematic analysis, we have tried several 
simple analytical ansatze for the K\"ahler potential as a function of $f^2$ which, together with our choice for
$W$, define the particle content. Within this context it is not difficult to find examples where all constraints 
are satisfied. As we will see in the next section, once we fix $(p, q)$, the boundary conditions at the origin
are essentially unique (modulo a rescaling of the coordinate $z$). Then we just have to evolve the different functions
from these regular boundary conditions at $r=0$ to the large $r$ regime. The problem is that, in most cases, the
solutions do not have a good asymptotic behaviour, i.e. the condition for localizing gravity in 4D is no longer 
fulfilled.

\subsection{A particular model}

A simple working example is described by
\begin{eqnarray}    
 K (\phi  \bar\phi) & = &  \frac{ \phi  \bar\phi}{1 - 10 \, \phi  \bar\phi}   \;,  \\
 W (\phi )         & = &  \phi   \;,  
\end{eqnarray}
which corresponds to choosing $p=1$, and gives the following  scalar potential
\be
V (\phi, \phib ) = 2 {\rm e}^{\left( \frac{\phi  \bar\phi}{1 - 10 \, \phi  \bar\phi}\right)}
\frac{2 - 81 \phi  \bar\phi + 1122  \phi^2  \bar\phi^2
 - 7100  \phi^3  \bar\phi^3 + 20000  \phi^4  \bar\phi^4} 
{1 - 100 \phi^2  \bar\phi^2} \;\;.
\ee
It is worth pointing out that the metric derived from this choice of K\"ahler potential becomes 
singular at $|\phi|^2 = 0.1 $ and is well defined only for smaller values of $|\phi|^2$. 
The potential is positive for small $|\phi|$ values and then decreases, entering a region where $V$ is 
negative. It has a minimum at $|\phi|^2 \sim 0.097 $, close to the barrier due to the above-mentioned singularity 
in the K\"ahler potential\footnote{ This singularity does not pose a problem since, as we will see, the string 
does not spread beyond such value. In fact we have found other, more involved, models where the  K\"ahler metric is 
well defined everywhere. For example, one can take the  K\"ahler potential presented here, perform a series expansion 
around the value $ |\phi|^2 =  0 $, and keep as many terms as necessary in order to reproduce the 
scalar potential shape given in fig.~\ref{fig:Vbueno} in the region of interest. 
Such potentials are perfectly well behaved, though one must keep the 
first eight terms of the K\"ahler expansion in order to reproduce all the features of the model.}. All these features 
are shown in fig.~\ref{fig:Vbueno}.
\FIGURE{\centerline{\includegraphics*[width=10cm, draft=false]{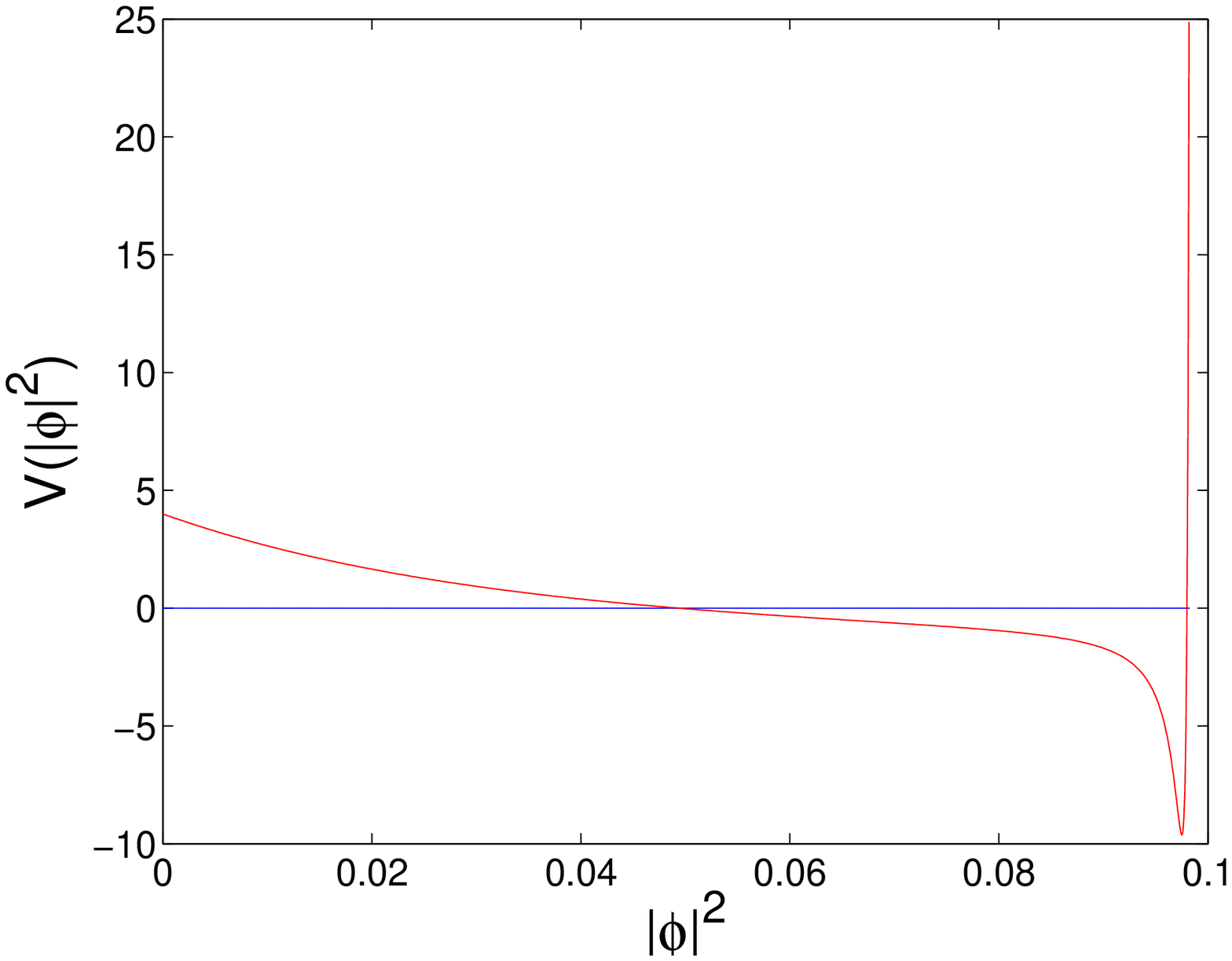}}
\caption{Plot of the scalar potential $V$ as a function of $|\phi|^2$ for the model described in the text.}
\label{fig:Vbueno}}
%
Let us now try to solve the BPS equations. Our ansatz for the scalar field and one of the warp factors is
\begin{eqnarray}
 \phi            & = &  f(r) {\rm e}^{i \theta} \;\;,\\
 {\rm e}^{N}           & = &  h(r)        \;\;.
\end{eqnarray}
Note that this choice corresponds to $q=1$ and $n=0$ ({\em i.e.}  $J$ does not depend on $\theta$). 
By solving the equations for $f_0$ and $h_0$, we get
\be
  f_0 = \frac{\sqrt{2}}{5}, \; h_0 = \frac{15}{\sqrt{2}} \;\;.
\ee
As we have mentioned before, $f_0^2 =0.08 < 0.1$ and, therefore, the asymptotic value for $|\phi|$ falls in 
the region where the model is well defined. Notice that $V <0 $  for $|\phi| = f_0$, which means that
the potential plays the role of a negative cosmological constant in the external region.

Concerning the string core, we have found that there is only one consistent set of regular boundary 
conditions\footnote{These boundary conditions apply only to the case $q=1$, however we have found that
there exists only one solution for any $q>0$, which is nothing but the one presented here expressed
in terms of the variable $z=u^q$.}.
This is given by $f \sim r$,  $h \sim const$. Since it is possible to rescale $z$ by redefining $A$, 
$B$ we can fix, at lowest order, $f = r$. We then get
\begin{eqnarray}    
f(r) & =  & r + {\cal O}(r^3) \;,
\\
h(r) & = & 1 +  \frac{r^2}{2} +  {\cal O}(r^3) \;,
\\
A(r) & =  &  - \frac{r^2}{2} +  {\cal O}(r^3) \;,
\\
B(r) & =   &  {\cal O}(r^4)  \;.
\end{eqnarray}
%
\FIGURE{\centerline{\includegraphics*[width=10cm, draft=false]{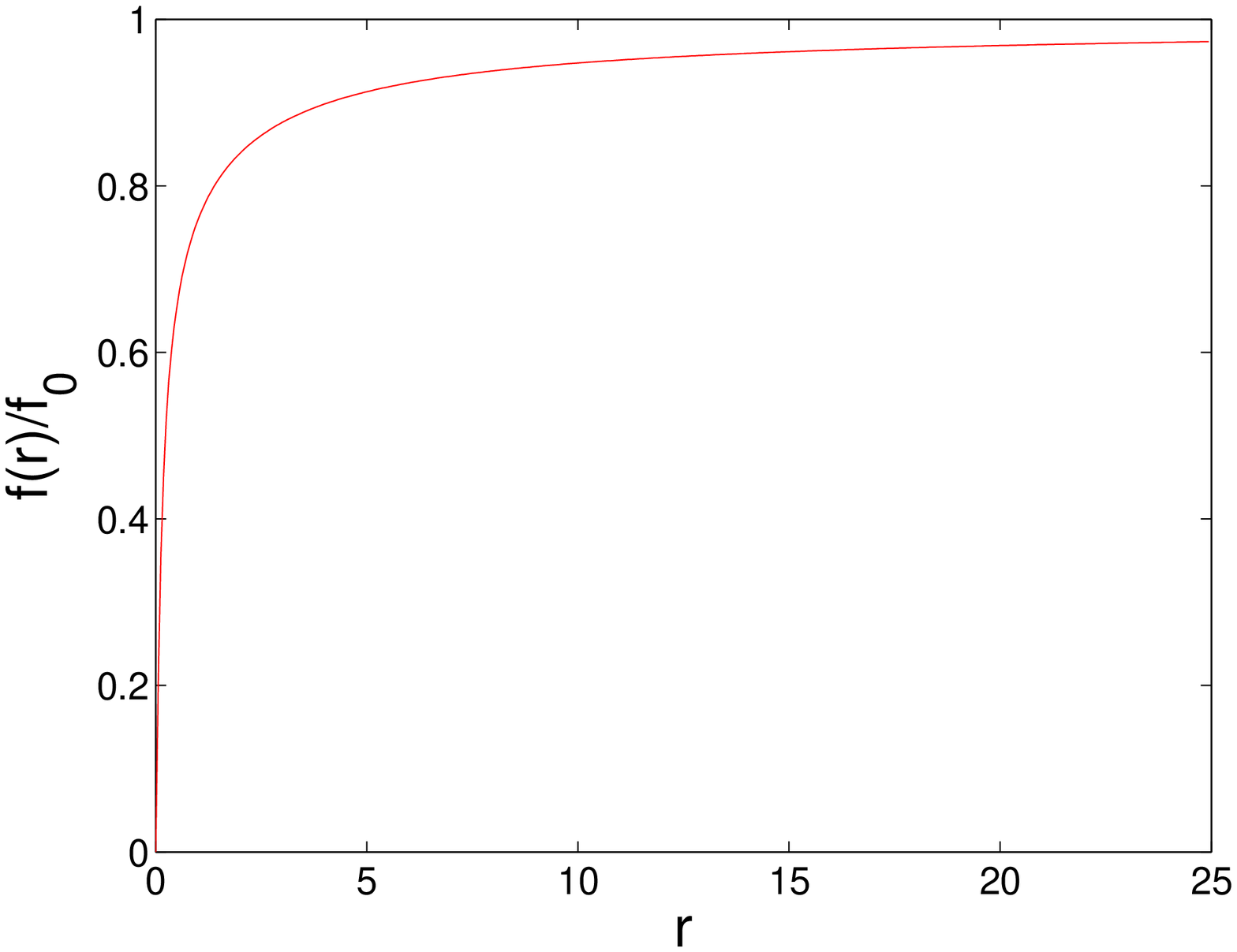}}
\caption{Plot of $f(r)/f_0$ as a function of $r$.}
\label{fig:efe}}
%

\FIGURE{\centerline{\includegraphics*[width=10cm, draft=false]{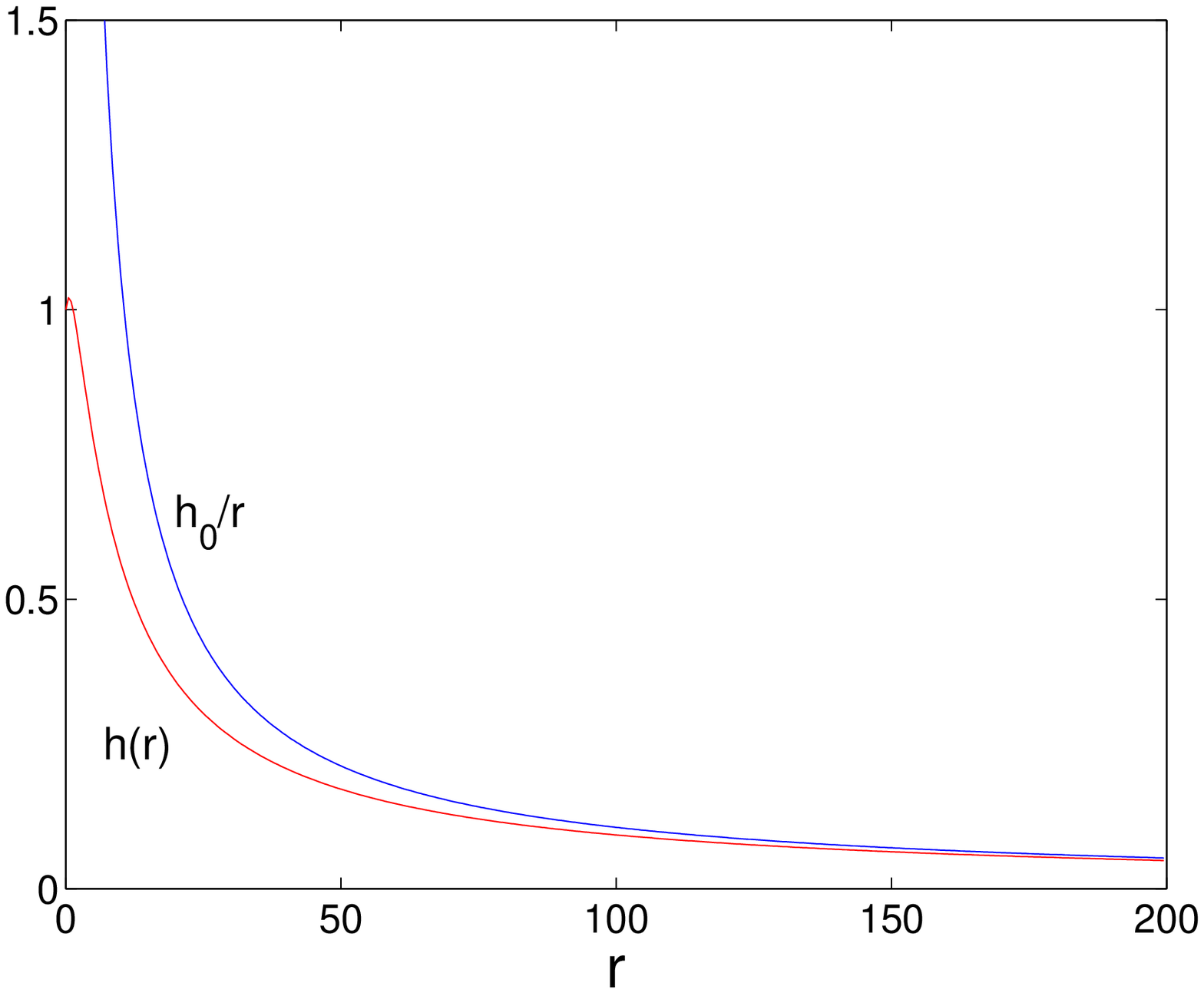}}
\caption{Plot of $h(r)$ and $h_0/r$ as a function of $r$.}
\label{fig:hache}}
Once we have fixed these boundary conditions, we just integrate the differential equations.
The result is shown in figs.~\ref{fig:efe}, \ref{fig:hache}.  As we anticipated, both functions approach their 
asymptotic values in the large $r$ region.  To illustrate this fact, we have plotted $f(r)/f_0$ in fig.~\ref{fig:efe}
and, in fig.~\ref{fig:hache}, we have added the theoretical asymptotic prediction for $h(r)$, namely $h_0/r$. Notice 
that the string core deduced from the scalar field stabilization, $f(r)$, is ${ \cal O} (10) $ smaller than the one 
given by the metric variation, $h(r)$.

Using eqs.~(\ref{N}), (\ref{defg}) and (\ref{defh}) we can evaluate the metric factors ${\rm e}^{2A(r)}$, 
${\rm e}^{2B(r)}$. These are shown in fig.~\ref{fig:eAeB}. For large $r$ values
\bea
{\rm e}^{2 A(r)} & \sim &  2 r^{-h_0 f_0^p } = r^{-6}  \;, \\
{\rm e}^{2 B(r)} & \sim &  \frac{h_0^2 \, {\rm e}^{- k[f_0^2]} } {r^2} = \frac{R_0^2}{r^2} \;.
\eea
%
\FIGURE{\centerline{\includegraphics*[width=10cm, draft=false]{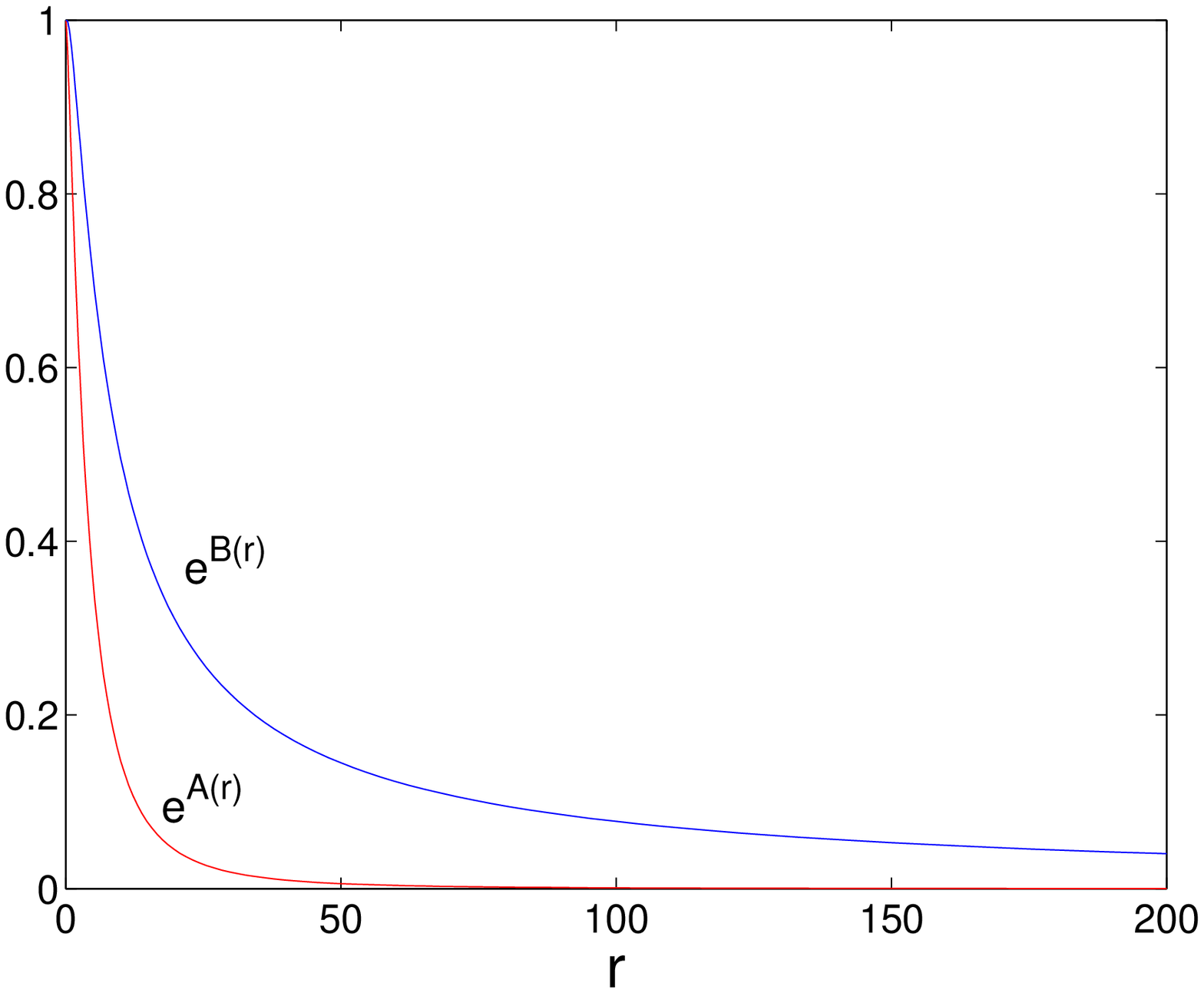}}
\caption{Plot of the two warp factors, ${\rm e}^{A(r)}$ and ${\rm e}^{B(r)}$ as a function of $r$.}
\label{fig:eAeB}}
%

Let us now compute the energy of the configuration described above. We will use the notation and results 
given in~\cite{Carroll:1999mu}. For static configurations, 
the energy stored in the transverse plane is defined as
\bea
E & = & E_{\rm grav} + E_{\rm kin} + E_{\rm pot}
\\
  & = & \int dx^5 dx^6\,  \sqrt{ g_\bot }\, \left(
- {1\over 2} R
+ K_{\phi \bar \phi } \, {\rm e}^{- 2 B} \frac{1}{2}
\left(  (\partial_{     z}  \phi) (\partial_{\bar z} \bar \phi) +
        (\partial_{\bar z}  \phi) (\partial_{z}      \bar \phi)        \right)
+ V(\phi, \bar \phi)
\right) \ ,
\nonumber
\label{energy}
\eea
and is equal to the negative of the action, as given by eq.~(\ref{lagrangian}). 
In ref.~\cite{Carroll:1999mu} is shown that it is possible to 
write the energy density as a sum of quadratic terms, each vanishing when the BPS equations are fulfilled, 
plus some total derivatives $ \Sigma_1, \Sigma_2, \Sigma_3$ given by
\bea
\Sigma_1 & = & \partial [ {\rm e}^{4 A} (8 \, \bar \partial A + 2  \, \bar \partial N -  \bar \partial K) ]
                + \; ( \partial \leftrightarrow  \bar \partial \; ,  N \leftrightarrow  N^* ) \, ,
\\[1mm]
\Sigma_2 & = & \partial [{\rm e}^{4 A} (K_\phi \bar \partial \phi - K_{\bar \phi}  \bar \partial \bar \phi )]
                + \; ( \partial \leftrightarrow  \bar \partial \; ,  \phi \leftrightarrow  \bar \phi ) \ ,
\\[1mm]
\Sigma_3 & = & 4 \partial [ {\rm e}^{4 A + N } \, W] + 4 \bar \partial [ {\rm e}^{4 A + N^*} \, \bar W]  \, . 
\eea
Then the energy of a BPS configuration is given by
\be
E = \int \, dx^5 \, dx^6 ( \Sigma_1 + \Sigma_2 +  \Sigma_3) \,  .
\label{toten}
\ee
We have found that it is still possible to simplify this expression, using the BPS equations, once again. We get
\be
E = \int \, dx^5 \, dx^6 \left(
2 \partial [ {\rm e}^{4 A} (\bar \partial A +  \, \bar \partial (N - N^*) ) ]
                + \; ( \partial \leftrightarrow  \bar \partial \; ,  N \leftrightarrow  N^* ) \right) \, .
\ee

Using Stokes' theorem 
\be
\int_R  dz \, \wedge d \bar z \, (\partial_z v_{\bar z} - \partial_{\bar z} v_{z})
=
\oint_{{\cal C} \equiv \delta R}  (dz \, v_z + d {\bar z} \, v_{\bar z}) \, ,
\ee
with
\be
v_{\bar z}  = \phantom{-} i \; 2 {\rm e}^{4A} \partial_{\bar z} (A + N - N^*) \, ,
\ee
we can calculate the energy stored in a region enclosed by a contour $\cal{C}$
\be 
  E_{\cal C}  =  - i \, \oint_{\cal C}
\left( 2 {\rm e}^{4A} \partial_{z} (A + N^*- N) dz - 2 {\rm e}^{4A} \partial_{\bar z} ( A + N - N^*) d \bar z
\right) \, .
\ee
Let us now aply this formula to an axially symmetric solution. The first thing to notice is that the contribution 
from $N$ cancels when this function is real, as happens to be the case in our example\footnote{
In general this term will vanish if the two real functions $(h, j)$ that parametrize $N$ have no singularities, 
as expected in regular solutions.}. In that situation the energy can be seen as uniquely dependent on the warp factor. 

Using that  
\be
z \, \partial_{z} \; {\rm e}^{4 A(z, {\bar z})}  =   r g(r) g'(r) \, , 
\ee
we can calculate the energy $E(r)$ stored in a circle of radius $r$ around the center of the string
\be    
 E (r)  =   2 \pi \,  r g(r) g'(r) \, .
\ee
We have verified this equation numerically. Taking the limit $ r \rightarrow \infty $  
we find that $E_{total} = 0$.

We should stress that this energy is zero for any  solution of that traps gravity. 
This follows from the regularity of the metric at the origin, $r=0$, and from the inverse power decay of 
the warp factor, $A(r)$. 
\FIGURE{\centerline{\includegraphics*[width=10cm, draft=false]{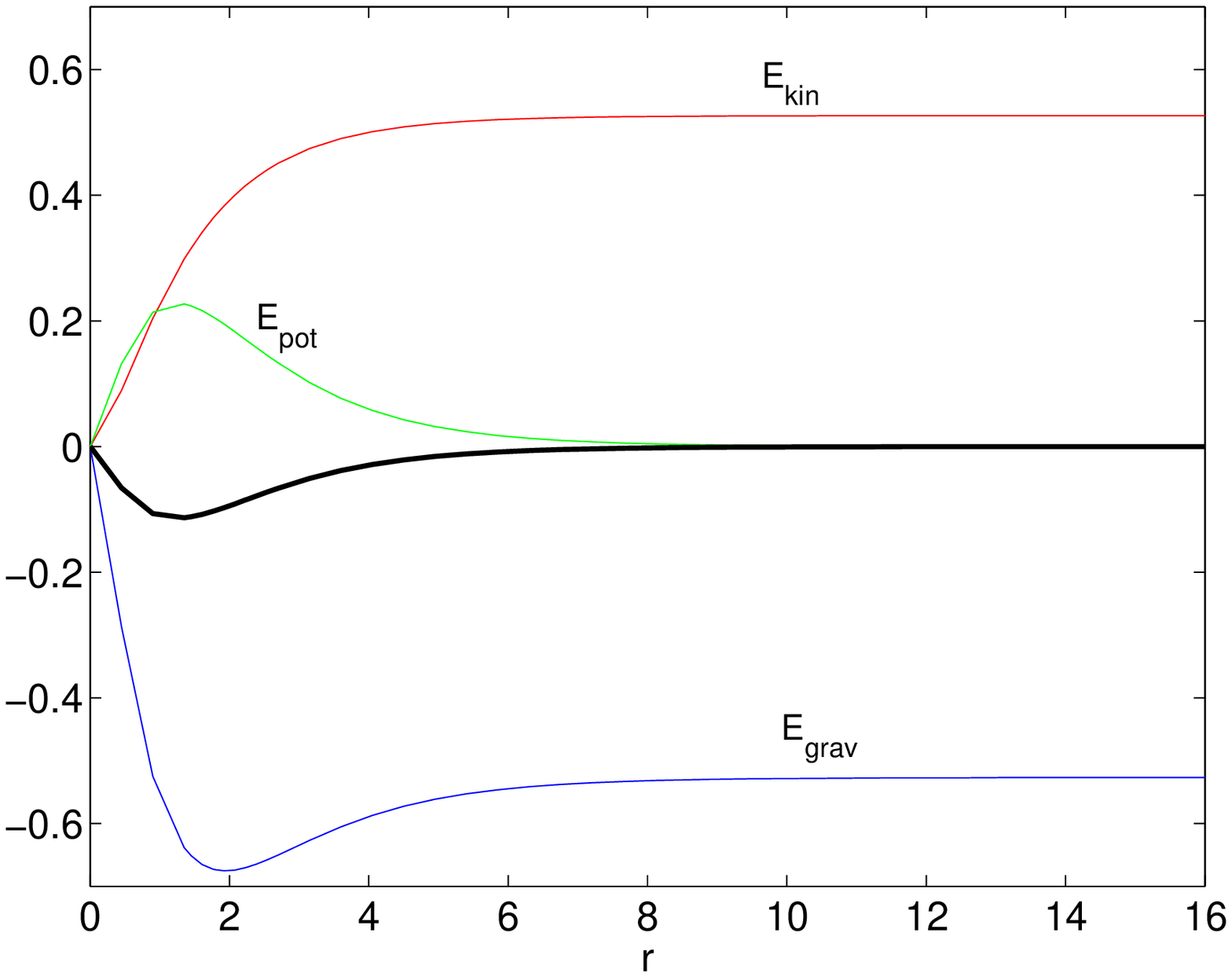}}
\caption{Plot of the energy enclosed in a circle of radius $r$ as a function of $r$. The 
total energy $E(r)$
is given by the thick black line, whereas its three different terms ($E_{\rm grav}(r)$ 
--blue--, $E_{kin}(r) $  --red--
and $E_{\rm pot}(r)$ --green--) are also shown.}
\label{fig:eners}}  
%
%
     
We illustrate the previous results in fig.~\ref{fig:eners}, where we have plotted $E(r)$, calculated by numerical 
integration, as a function of $r$ (thick black line). We have also plotted the three different contributions to 
the energy, coming from the integration of  the curvature (blue solid line), kinetic (red) and potential 
(green) energy densities as given in (\ref{energy}). 
A remarkable fact is that the total (integrated) potential energy goes to zero. This can be easily 
understood by calculating the energy of a family of configurations that are just scaling deformations of the solution, 
and using the fact that our defect spreads along two spatial dimensions. The vanishing of the total potential energy 
sets stringent bounds on the class of models of gravity plus scalar field that could work. In particular, models with 
canonical kinetic terms, as presented in section~3, are ruled out, as it is now obvious from fig.~\ref{fig:Vp}, where 
the potential happens to be negative in the region where the string would spread.

\section{Discussion}

In this section we would like to elaborate further on the main results achieved. In order to do that, let us
present first the geometry that results of our solution. To see it more straightforwardly, it is useful to
realize that there are other parametrizations of metrics in six dimensions with axial symmetry. 
A popular one is given by a normalized radial contribution to the metric, i.e
\be
ds^2 =  \sigma^2 (\rho) \, \eta_{\mu\nu}dx^\mu dx^\nu + d\rho^2 + \gamma^2(\rho) \, 
d\theta^2 \;. 
\ee
We have expressed our solutions in terms of these coordinates. The change of variables is regular everywhere. 
The shape of the $\sigma (\rho)$ and $\gamma (\rho)$ factors is drawn in fig.~\ref{fig:sigam}.
\FIGURE{\centerline{\includegraphics*[width=10cm, draft=false]{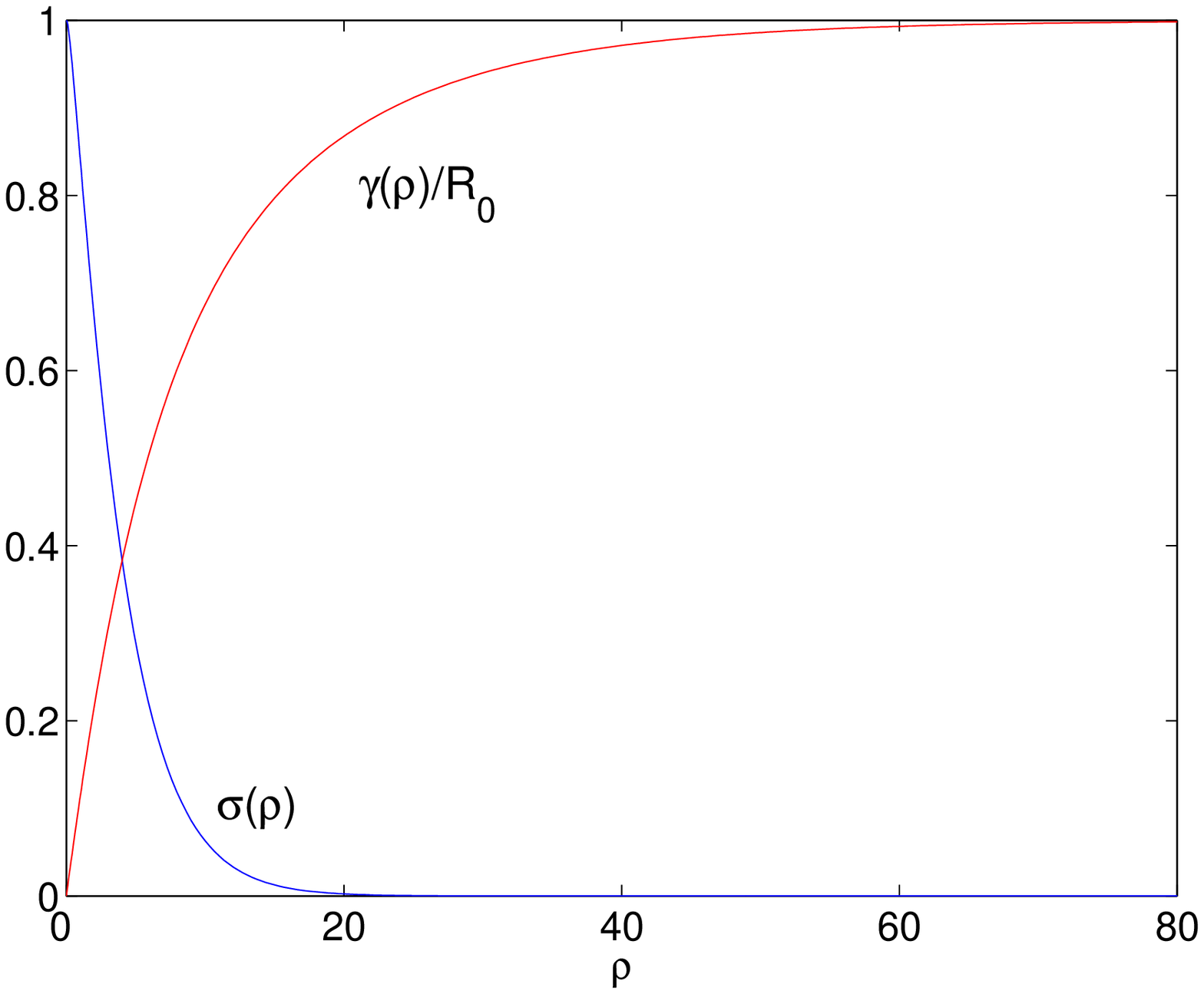}}
\caption{$\sigma(\rho)$ and $\gamma(\rho)/R_0$ as a function of $\rho$ for the solution presented in the previous 
section.}
\label{fig:sigam}}
Far from the core, the  $\sigma$ warp factor decreases exponentially
\be
 \sigma^2 (\rho) \sim  {\rm e}^{- \rho \, \alpha } \; , 
\ee
with
\be
    \alpha^2 =  - \frac {V[f_0^2]}{2} \;\;, 
\label{alpha}
\ee
and  $\gamma$ goes to a constant given by
\be
 R_0 =  |h_0|  \, {\rm e}^{-k[f_0^2]/2} \; .
\ee
The space factorizes as  $AdS_5 \times S_1 $ and, therefore, the curvature in this region is constant. This geometry, 
with a cylindrical transverse space and an exponential warp factor, was proposed  by Gregory~\cite{Gregory:1999gv} 
as the one allowing for nonsingular global string compactifications in D=6, and also by Olasagasti and 
Vilenkin~\cite{Olasagasti:2000gx} in their generalization to higher dimensional global defects.

Inside the core, $\gamma (\rho) \sim \rho $ and, consequently, there is no deficit angle. This follows from our 
choice of boundary conditions at the origin, as can be seen using that $ \rho \sim r$ for small values of $\rho$. 

Our solution is regular everywhere and interpolates smoothly between the origin and the asymptotic region, providing 
a cigar-like, complete geometry. As we mentioned before, this kind of geometry was proposed by Gregory in 
ref.~\cite{Gregory:1999gv}. She considered a global string given by a potential with $V_{\min}$ = 0 and a negative 
cosmological constant and argued that, for some values of the parameters, it should be possible to match the two 
regions building a nonsingular metric. We have presented here an example based on this idea, although there is one 
difference which is important enough to be pointed out. In our example there is no cosmological constant. The required 
negative contribution to the energy density far from the core is provided by the potential and, therefore, depends 
on the asymptotic value for $ |\phi|$. This value is not necessarily, as one could naively think, the one that 
minimizes the potential. It is dynamically fixed by the field equations which implies that, in general, we can not 
neglect the feedback of the metric on the stabilization of the modulus of the scalar field. 
\FIGURE{\centerline{\includegraphics*[width=10cm, draft=false]{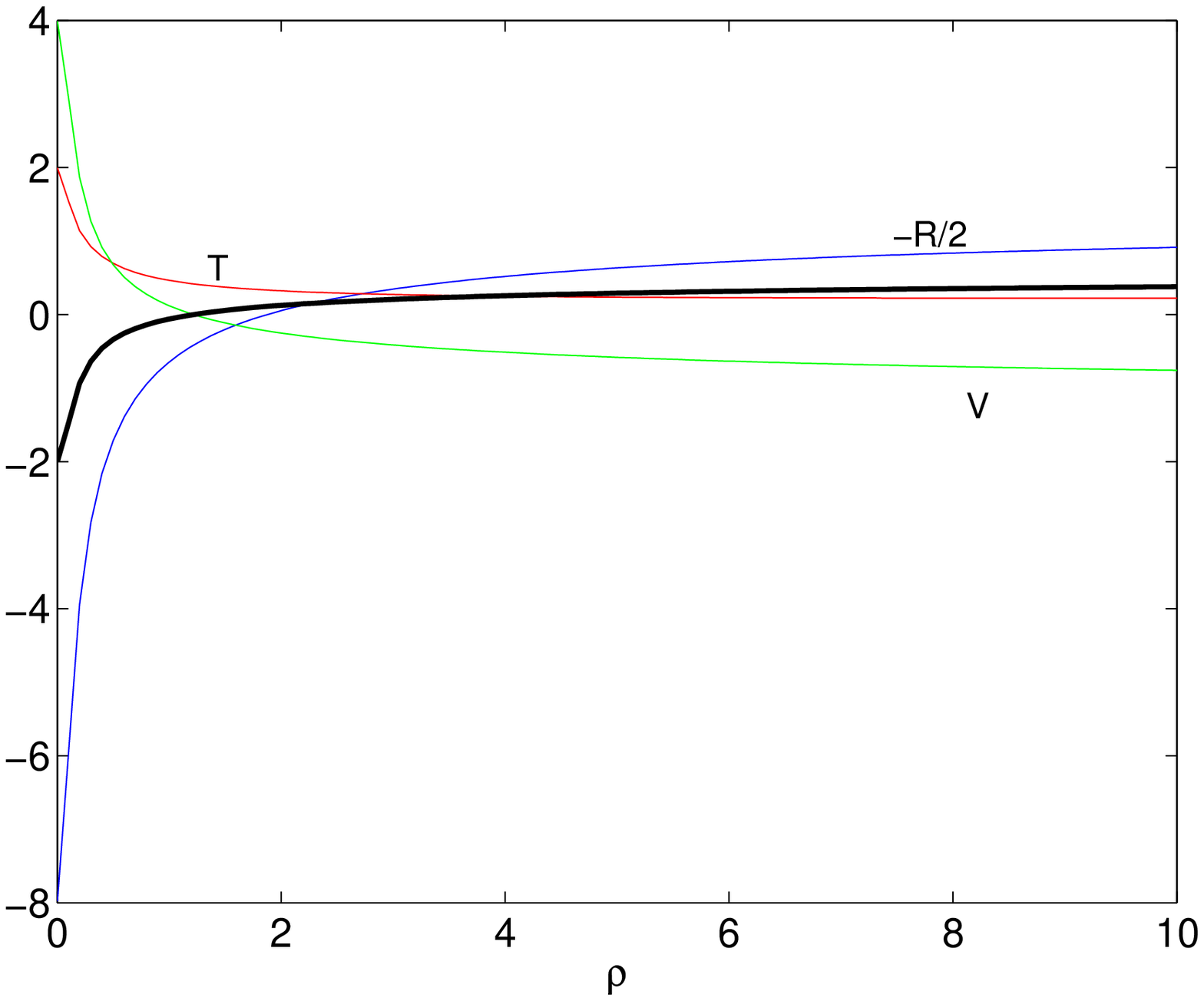}}
\caption{Plot of the different energy densities as a function of $\rho$. As in fig.~\ref{fig:eners}, here the total
energy density is given by the thick black line, whereas its different contributions are given by the blue (curvature),
red (kinetic) and green (potential) lines.}
\label{fig:densis}}
%
In fig.~\ref{fig:densis} we plot the gravitational, kinetic and potential energy densities, as well as their sum. 
The pattern shown in the figure is quite general. As explained before, far from the core the value of the curvature 
is fixed by the scalar --winding plus potential-- energy, $\Lambda_{\phi}$, through the  $AdS$ relation
\be
R = \frac{2 d}{d-2} \Lambda_{\phi} \, 
\ee
whith $d = 5$. We can check that the negative potential energy density dominates over the kinetic one in this 
asymptotic region. Since the energy coming from the integration of the curvature term has to cancel the total 
kinetic energy, the curvature must be positive in the inner region, as shown in  fig.~\ref{fig:densis}.
The value of $\rho$ at which the scalar curvature vanishes ($\rho = 2$) is of the same order of magnitude than 
the one where $V$ also does ($\rho = 1.3$).

The previous discussion leads to the next issue we want to address, namely the phenomenological applications 
of these models. From the features of the energy configurations cited above, it turns out that we can
observe how the relevant parameters of the model (i.e. the two warp factors and the modulus of the scalar field)
adjust themselves to their asymptotic values, $\alpha$, $R_0$ and $f_0$, in order to fulfill them. In fact,
we can relate the asymptotic value of $\alpha$ to the value of the scalar potential at $f_0$, $V[f_0^2]$,
as we have written in eq.~(\ref{alpha}).

This means, among other things, that obtaining the right hierarchy between the D=6 and D=4 Planck masses
(mainly parameterized by $\alpha$), will crucially depend on the dynamics of the scalar field, namely
its K\"ahler potential and its superpotential. Indeed the trapping condition eq.~(\ref{trapping}) 
has to be verified in order to have a finite four-dimensional reduced Planck mass. Also the transverse volume has 
to be large enough to generate the appropriate value for the Planck mass from the six-dimensional parameters
(these are the higher-dimensional scale of gravity, $M$, and the parameters upon which $W$ and $K$
depend). In the case we have presented this volume (see eq.~(\ref{criterio})) is $O(100)$, therefore 
a gravity scale hierarchy between D=6 and D=4 is not generated.

We have not yet estimated the correction to Newton's law in this kind
of scenarios. 
In general we expect a mixture of effects that typically 
appear in the case of one extra transverse dimension plus some Kaluza-Klein
ones modulated by the presence of the warp factor. 
Since we have a complete description of the metric, 
one could perform, at least numerically, a complete analysis 
within a realistic model. Meanwhile, we can use the results of two
scenarios that share some of the features of our model. 
In ref.~\cite{Kogan:2001yr}, a single, positive tension four-brane embedded in a six-dimensional
$AdS$ space is considered. One transverse dimension is compactified and the 
other is taken to be an infinite one. They consider axially symmetric, constant energy momentum tensors. 
These bulk and brane tensors are allowed to be inhomogeneous and include, in
particular, the typical pattern of asymptotic string-like defects \cite{Gherghetta:2000qi}.
The flatness of the brane and the specific cylindrical geometry are obtained
by several fine tunings among the bulk and brane energy momentum tensor components.
The correction  to Newton's law in this scenario is shown to be small.
Another particular case, providing this kind of energy momentum tensor, 
is obtained by considering an Abelian magnetic field 
\cite{Gibbons:1986wg, Gherghetta:2000jf}. The radius of the
cylinder is fixed by the magnetic charge and by the bulk cosmological constant.

We want to stress that in these models there is a core where 
the metric is unknown since the asymptotic solution is not valid there.
On the other hand, in order to evaluate the corrections to Newton's
law, it is either necessary to make some assumptions on the form
of the metric near $\rho = 0$, or one has to work in the thin-core limit. 
Since this latter approach is not always justified, our solution could
provide a model to explore general thick core corrections to these
approximations.

There are other issues that one has to study when the core of the defect 
is replaced by a brane with some tension. In D=5, for 
example, one is almost forced to include negative tension branes
to generate decreasing warp factors. This is dangerous, since
the energy conditions will, in general, be violated. 
Also one has impose energy momentum conservation by hand, getting relations
between brane tensions and metric discontinuities. 
Since we are dealing with a real topological defect, 
our model does not suffer this kind of anomalies. The
metric is regular everywhere and the energy momentum tensor 
is conserved -- since it is derived from a Lagrangian
density --. Also the  null energy condition
\be
T_{\alpha \beta }v^\alpha v^\beta \geq 0 \;\;,
\ee
with $v^\alpha$ a null vector, is automatically fulfilled since
\be
T_{\alpha \beta } v^\alpha v^\beta  =  2 |v^z|^2
  \; K_{\phi {\bar \phi}}
  \left( 
          (\partial        \phi) ({\bar \partial} {\bar \phi} ) +
          ({\bar \partial}  \phi) (\partial        {\bar \phi} )
  \right) 
\ee
is always positive.

In conclusion we have presented here a complete numerical solution to the system of Einstein and field equations 
of a six-dimensional setup with gravity and a scalar field. This was achieved by solving a particular set of
first order differential equations, the BPS equations, whose solutions also satisfy the more general second order 
system. Using analytic ansatze on the behaviour of the field and warp factors of the metric around the
origin of the defect and in the aymptotic region, we have been able to classify the solutions obtained and
determine what sort of dynamics (i.e. K\"ahler potential and superpotential) must the scalar field obey in order 
to have gravity trapping around the four-dimensional defect. The geometry of the phenomenologically
viable solutions is cigar-shaped and those configurations have a total vanishing energy.

\acknowledgments

We thank M. Giovannini, R. Gregory and T. Ort\'{\i}n for answering all our queries. BdC thanks the CERN Theory Division
and IEM (CSIC) for hospitality and financial support during different stages of this work. JMM thanks the University 
of Sussex for its hospitality and the Royal Society for support through the award of a study visit.

\end{document}